\documentclass[letterpaper,twocolumn,english,aps,prb,showpacs,superscriptaddress]{revtex4-1}
\usepackage[T1]{fontenc}
\usepackage[latin9]{inputenc}
\usepackage{color}
\usepackage{amsmath}
\usepackage{amssymb}
\usepackage{graphicx}

\makeatletter

\usepackage{ulem}
\usepackage{float}
\usepackage{lipsum}
\usepackage{babel}

\makeatother

\usepackage{babel}
\begin{document}
\title{Exact zeros of the Loschmidt echo and quantum speed limit time for the dynamical
quantum phase transition in finite-size systems}
\author{Bozhen Zhou}
\affiliation{Beijing National Laboratory for Condensed Matter Physics, Institute
of Physics, Chinese Academy of Sciences, Beijing 100190, China}
\affiliation{School of Physical Sciences, University of Chinese Academy of Sciences,
Beijing 100049, China }
\author{Yumeng Zeng}
\affiliation{Beijing National Laboratory for Condensed Matter Physics, Institute
of Physics, Chinese Academy of Sciences, Beijing 100190, China}
\affiliation{School of Physical Sciences, University of Chinese Academy of Sciences,
Beijing 100049, China }
\author{Shu Chen}
\email{Corresponding author: schen@iphy.ac.cn }

\affiliation{Beijing National Laboratory for Condensed Matter Physics, Institute
of Physics, Chinese Academy of Sciences, Beijing 100190, China}
\affiliation{School of Physical Sciences, University of Chinese Academy of Sciences,
Beijing 100049, China }
\affiliation{Yangtze River Delta Physics Research Center, Liyang, Jiangsu 213300,
China }
\date{\today}
\begin{abstract}
We study exact zeros of Loschmidt echo and quantum speed limit time
for dynamical quantum phase transition in finite size systems. Our
results illustrate that exact zeros of Loschmidt echo exist even in
finite size quantum systems when the postquench parameter takes some
discrete values in regions with the corresponding equilibrium phase
different from the initial phase. As the system size increases and
tends to infinity, the discrete parameters distribute continuously
in the parameter regions. We further analyze the time for the appearance of the first exact
zero of Loschmidt echo which is known as the quantum speed limit time $\tau_{\text{QSL}}$. We demonstrate that
the maximal value of $\tau_{\text{QSL}}$ is proportional to $L$
and approaches infinity in the thermodynamical limit, when we quench
the initial non-critical state to the critical phase. We also calculate
the minimal value of $\tau_{\text{QSL}}$ and find that its behavior
is dependent on the phase of initial state.
\end{abstract}
\maketitle

\section{Introduction}

In recent years, the development of quantum simulation platforms,
such as neutral atom arrays\citep{Bernien2017Nature,Omran2019Science,Bluvstein2021Arxiv,Ebadi2021Arxiv},
stimulates the intensive studies on the nonequilibrium dynamics of
quantum many-body systems. An interesting issue is the dynamical quantum
phase transition (DQPT) {\citep{Heyl2013PRL,Karrasch2013PRB,Canovi2014PRL,Andraschko2014PRB,Marcuzzi2014PRL,
Hickey2014PRB,Heyl2014PRL,Heyl2015PRL,Vajna2015PRB,Schmitt2015PRB,Abeling2016PRB,Budich2016PRB,
Zvyagin2016LTP,Heyl2018RPP,Yang2017PRB,Heyl2018PRL,Mera2018PRB,Jurcevic2017PRL,Guo2019PRAp,
Flaschner2018Nature,Wang2019PRL,BoBo2020PRL,BoBo2020PRB,Halimeh,Jafari,Halimeh2,Jafari2,Pedersen2021PRB,YangC}, which describes dynamical quantum critical phenomena presented in
quench dynamics of kinds of quantum systems with initial state chosen
as the ground state of a given Hamiltonian and evolving under a sudden
change of a Hamiltonian parameter. The DQPT is characterized by the
emergence of zero points of Loschmidt echo (LE) at a series of critical
times, where the LE is defined by $\mathcal{L}(t)=\left|\mathcal{G}(t)\right|^{2}$
with the Loschmidt amplitude given by
\begin{equation}
\mathcal{G}(t)=\langle\psi_{i}|e^{-iH_{f}t}|\psi_{i}\rangle.
\end{equation}
Here $|\psi_{i}\rangle$ is the ground state of prequench Hamiltonian,
$H_{f}$ is the postquench Hamiltonian, and we have set $\hbar=1$.
The LE measures the overlap between initial quantum state and time-evolved
state\citep{Gorin2006PR}, which has wide application in various contexts
ranging from the theory of quantum chaos\citep{Gorin2006PR,Peres1984PRA,Jalabert2001PRL}
and the Schwinger mechanism of particle production\citep{Martinez2016Nature,Schwinger1951PR},
to work distribution functions in the context of nonequilibrium fluctuation
theorems\citep{Talkner2007PRE,Palmai2015PRB}. The existence of zero
points of LE means the occurrence of nonanalytic behaviors of dynamical
free energy, i.e., the rate function of LE given by $\lambda(t)=-\frac{1}{L}\ln\mathcal{L}(t)$,
at these critical times. It has been shown that DQPT and the equilibrium
quantum phase transition are closely related as the nonanalyticities
in the rate function of LE occur for quenches crossing the static
quantum phase transition point\citep{Heyl2013PRL,Karrasch2013PRB,Hickey2014PRB},
although a one-to-one correspondence between them does not always
hold true\citep{Andraschko2014PRB,Schmitt2015PRB,Vajna2014PRB,Sharma2015PRB,Jafari2019SciRep}.
The relation between the long-time average of the LE and the ground
state fidelity susceptibility\citep{Zanardi2006PRE,You2007PRE,Chen2008PRA,Chen2007PRE,Zhou2008JPA}
was also unveiled recently\citep{Zhou2019PRB}.

In general, exact zeros of LE or nonanalyticities of dynamical free-energy
only occur when the system size tends to infinity. This is very similar
to Fisher zeros of the partition function in statistical physics\citep{Fisher1965}.
It is well known that the Fisher zeros in a finite size system do
not lie on the real temperature axis, and exact zeros only emerge
in the thermodynamic limit \citep{Fisher1965,Yang1952PR,Saarloos1984JPA}.
Similarly, the exact zeros of LE in a finite size quantum system can
only appear in the complex time plane. When the system size tends
to infinity, the zeros approach to the real time axis for quenches
crossing the quantum phase transition point\citep{Heyl2013PRL}. Now
a question arise here, one may ask whether exact zeros of LE can occur
in real time axis for a finite size quantum system? If the answer
is yes, it seems that there exists controversy with the known results
and how to understand the seeming controversy?

Aiming to answer the above questions and deepen our understanding
of DQPT in the finite size systems, we shall explore the exact zeros
of LE by focusing on a well-known exact solvable model, i.e., the
transverse field Ising model (TFIM), which is well studied and known
to exhibit DQPT in the thermodynamic limit. The existence of exact
solutions enable us to analytically derive the condition for the
existence of exact zeros of LE in finite size systems. For a given
initial state prepared as the ground state of pre-quench Hamiltonian,
our results illustrate that exact zeros of LE exist even in finite
size quantum systems when the post-quench parameter takes some discrete
values. These discrete parameters are found to locate in regions with
the corresponding equilibrium phase different from the initial phase.
As the system size increases and tends to infinity, the discrete parameters
distribute continuously in the parameter regions and thus are consistent
with previous results.

Further, once we know the exact zeros of LE in finite size quantum
systems, it is natural to explore the minimum
time of an initial state evolving to its orthogonal state which
corresponds to the time for the emergence of the first exact
zero of LE. The minimum time required for
arriving an orthogonal quantum state is called quantum speed limit
(QSL) time , denoted as $\tau_{\text{QSL}}$. The
QSL time gives fundamental limit on the time scale for how fast a
quantum state evolves in real-time dynamics, and the lower bound of  $\tau_{\text{QSL}}$
has been discussed in closed quantum systems \citep{Fleming1973,Bhat1983JPAMG,Anandan1990PRL,Vaidman1992AJP,Uffink1993AJP,Pfeifer1993PRL,Margolus1998Physica,Giovannetti2003PRA,Levitin2009PRL,Fogarty2020PRL,Nikolai2021PRA}
and open quantum sytems \citep{Campo2013PRL,Deffiner2013PRL,Mirkin2016PRA,Vu2021PRL}.
The discussion of QSL time can be traced back to the early time when
Mandelstarn and Tamm studied the time-energy uncertainty in non-relativistic
quantum mechanics \citep{MT1945JP}. The QSL time is also related
to several interesting topics, such as quantum optimal control \citep{Frank14187SciRep},
quantum information\citep{Pires2020PRR} and quantum geometry\citep{Pires2016PRX}.
In the framework of DQPT, the QSL time has been studied in a previous
work \citep{Heyl2017PRB}, which however, only concerns on the dynamics
of the quantum critical state. As we shall clarify in this work, if the initial system is in the quantum critical state, no exact zeros of LE can be found  and thus the connection of QSL time to the DQPT is still elusive. In this work, we shall unveil how the QSL time changes with
quench parameters and explore the general connection between QSL time and DQPT by considering
various situations with different quench parameters.
We also pay particular attention to the
maximum and minimum value of $\tau_{\text{QSL}}$, as the maximum of $\tau_{\text{QSL}}$ is related to the quench dynamics close to critical point and the minimum of $\tau_{\text{QSL}}$ gives the important message for how fast the DQPT could happen. When the system size tends to infinity, we find that the maximum value of $\tau_{\text{QSL}}$ approaches
infinity when the quench parameter approaches to the critical point, which is independent of the initial state. However, the behavior
of the minimum value of $\tau_{\text{QSL}}$ is distinct if the initial
state is chosen in different phase. We demonstrate that non-analytical
behaviours appear in both the average of $\tau_{min}(L)$ and the
variance of $\tau_{min}(L)$ when we change the prequench parameter
across the static quantum phase transition point.

The article is organized as follows. In Sec. II, we study the quench dynamics in finite size systems of the transverse field Ising model and give exactly the relation for the occurrence of zeros of LE. The dynamical behavior quenched to the parameter region close to the critical point is also studied. In Sec. III, we study quantum speed limit
time under different quench parameters and explore its connection to DQPT. A summary and outlook is given in Sec.IV.

\section{Exact zeros of Loschmidt echo in finite size systems}

We consider the one-dimensional (1D) 
TFIM described by the following Hamiltonian \citep{Pfeuty1970AP}:
\begin{eqnarray}
H & = & -J\sum_{j=1}^{L}\sigma_{j}^{x}\sigma_{j+1}^{x}-h\sum_{j=1}^{L}\sigma_{j}^{z},\label{eq:HIsing}
\end{eqnarray}
where $J$ is the nearest-neighbor spin coupling, $h$ is the external
magnetic field along the $z$ axis and the periodical boundary condition
$\sigma_{L+1}^{x}=\sigma_{1}^{x}$ is assumed. The three Pauli matrices
are $\sigma_{j}^{\alpha}(\alpha=x,y,z)$, $j=1,\cdots,L$ with $L$
denoting total number of lattice sites. The TFIM fulfills a duality
relation \citep{Fradkin1978PRD,Zhang2019PRL}:
$
UH(J,h)U^{-1}=JhH(1/J,1/h) .
$
By using the Jordan-Wigner transformation, the even-parity and odd-parity
of the TFIM with periodical boundary condition can be mapped to the
anti-periodical Kitaev chain and periodical Kitaev chain, respectively
\citep{Mbeng2009Arxiv,Lieb1961}. Then we can write the Hamiltonian
in the fermion representation as
\begin{align}
H= & -J\sum_{j=1}^{L-1}\left(c_{j}^{\dagger}c_{j+1}+c_{j}^{\dagger}c_{j+1}^{\dagger}+\text{H.c.}\right)-2h\sum_{j=1}^{L}c_{j}^{\dagger}c_{j}\nonumber \\
 & \pm J(c_{L}^{\dagger}c_{1}+c_{L}^{\dagger}c_{1}^{\dagger}+\text{H.c.}),\label{H-JW}
\end{align}
where the plus sign or minus sign is corresponds to the even-parity
or odd-parity. For convenience, we take $J>0$ in the following
discussions so that the system is in the ferromagnetic phase when
$|h/J|<1$.

It is convenient to diagonalize the Hamiltonian (\ref{H-JW}) in the
momentum space by using the Fourier transform $c_{j}^{\dagger}=\frac{1}{\sqrt{L}}\sum_{k}e^{ikj}c_{k}^{\dagger}$.
Here values of $k$ should be chosen in the set of $\mathcal{K}_{\text{PBC}}=\left\{ k=\frac{2\pi m}{L}|m=-L/2+1,\cdots,0,\cdots,L/2\right\} $
for periodical boundary condition (PBC) and $\mathcal{K}_{\text{aPBC}}=\left\{ k=\pm\frac{\pi(2m-1)}{L}|m=1,\cdots,L/2\right\} $
for anti-periodical boundary condition (aPBC) \citep{Mbeng2009Arxiv,Damski}.
In the following discussion, we focus on the even site of lattice
with even parity which is corresponding to aPBC. It should be noted
that all terms of Hamiltonian come into pairs $(k,-k)$ for aPBC.
Define the positive $k$ values as $\mathcal{K}_{\text{aPBC}}^{+}=\left\{ k=\frac{\pi(2m-1)}{L}|m=1,\cdots,L/2\right\} $.
Then the Hamiltonian in momentum space is
\begin{align}
H= & -2\sum_{k\in\mathcal{K}_{\text{aPBC}}^{+}}\left[\left(J\cos k+h\right)\left(c_{k}^{\dagger}c_{k}-c_{-k}c_{-k}^{\dagger}\right)\right.\nonumber \\
 & \left.-\left(iJ\sin kc_{k}^{\dagger}c_{-k}^{\dagger}+\text{H.c.}\right)\right].
\end{align}
By using the Bogoliubov transformation
\begin{eqnarray*}
\beta_{k} & =\cos\theta_{k}c_{k}+i\sin\theta_{k}c_{-k}^{\dagger},\\
\beta_{-k}^{\dagger} & =i\sin\theta_{k}c_{k}+\cos\theta_{k}c_{-k}^{\dagger},
\end{eqnarray*}
where $\frac{\epsilon_{k}}{E_{k}}=\cos2\theta_{k}$ and $\frac{\zeta_{k}}{E_{k}}=\sin2\theta_{k}$
with $\epsilon_{k}=-J\cos k-h$ and $\zeta_{k}=-J\sin k$, we arrive
at a Hamiltonian given by \citep{Pfeuty1970AP,Lieb1961,Mbeng2009Arxiv}
\begin{equation}
H=2\sum_{k\in\mathcal{K}_{\text{aPBC}}^{+}}\left(E_{k}\beta_{k}^{\dagger}\beta_{k}-E_{k}\beta_{-k}\beta_{-k}^{\dagger}\right),
\end{equation}
where $E_{k}=\sqrt{\epsilon_{k}^{2}+\zeta_{k}^{2}}$.

Then we consider the quench dynamics driven by the transverse field
$h$ which can be described by $h(t)=h_{i}\Theta(-t)+h_{f}\Theta(t)$.
The analytical formula of LE has the form
\begin{equation}
\mathcal{L}(t)=\prod_{k\in\mathcal{K}_{\text{aPBC}}^{+}}\left[1-\sin^{2}(2\delta\theta_{k})\sin^{2}(2E_{kf}t)\right],\label{eq:anaLE_Ising}
\end{equation}
where $\delta\theta_{k}=\theta_{kf}-\theta_{ki}$, 
\[
\theta_{ki}=\frac{1}{2}\arctan\frac{J\sin k}{J\cos k+h_{i}}
\]
is the Bogoliubov angle of prequench Hamiltonian, 
\[
\theta_{kf}=\frac{1}{2}\arctan\frac{J\sin k}{J\cos k+h_{f}}
\]
is the Bogoliubov angle of the postquench Hamiltonian, and $E_{kf}$ is
the energy of the postquench Hamiltonian. To ensure $\mathcal{L}(t)=0$,
we must have $\sin^{2}(2\delta\theta_{k})=1$, which gives rise to
the following constraint relation
\begin{equation}
\frac{h_{f}}{J}=-\frac{J+h_{i}\cos k}{h_{i}+J\cos k}.\label{eq:Ising_hihf_k}
\end{equation}

For a finite size system, the momentum $k$ takes discrete values.
Given the prequench parameter $h_{i}$, we can get a series of $h_{f}$
determined by Eq. (\ref{eq:Ising_hihf_k}) for various $k$. When the
postquench parameter takes these discrete values, we have $\mathcal{L}(t)=0$
at
\begin{equation}
t=t_{n}^{*}=\frac{\pi}{2E_{kf}}(n+\frac{1}{2}),
\end{equation}
with
\begin{equation}
E_{kf}/J=\sqrt{(\cos k+h_{f}/J)^{2}+\sin^{2}k},\label{Ekf}
\end{equation}
i.e., there exist exact zeros of LE as long as Eq. (\ref{eq:Ising_hihf_k})
is fulfilled. According to Eq. (\ref{eq:Ising_hihf_k}),\textcolor{red}{{}
}if $h_{i}/J\in(-1,1)$, the exact zeros of LE emerge only for\textcolor{red}{{}
}$h_{f}/J\in(-\infty,-1)\cup(1,\infty)$\textcolor{red}{. }For the
1D transverse field Ising chain, we can prove that the Loschmidt echo
fulfills the following dynamical duality relation
\begin{equation}
\mathcal{L}(\gamma_{i},\gamma_{f},t)=\mathcal{L}(\gamma_{i}^{-1},\gamma_{f}^{-1},\gamma_{f}t),\label{eq:ddr}
\end{equation}
where $\gamma_{i}=h_{i}/J$ and $\gamma_{f}=h_{f}/J$ are the dimensionless
parameters. Due to the existence of dynamical duality relation (see
appendix for details), we only need to consider the case of $h_{i}/J\in(-1,1)$\textcolor{red}{{}
}as the cases of $h_{i}/J\in(-\infty,-1)$ and $h_{i}/J\in(1,\infty)$
can be obtained by using the dynamical duality relation.
%%%%%%%%%%%%%%%%%%%%%%%%%%%%%
\begin{figure}[h]
\begin{centering}
\includegraphics[scale=0.24]{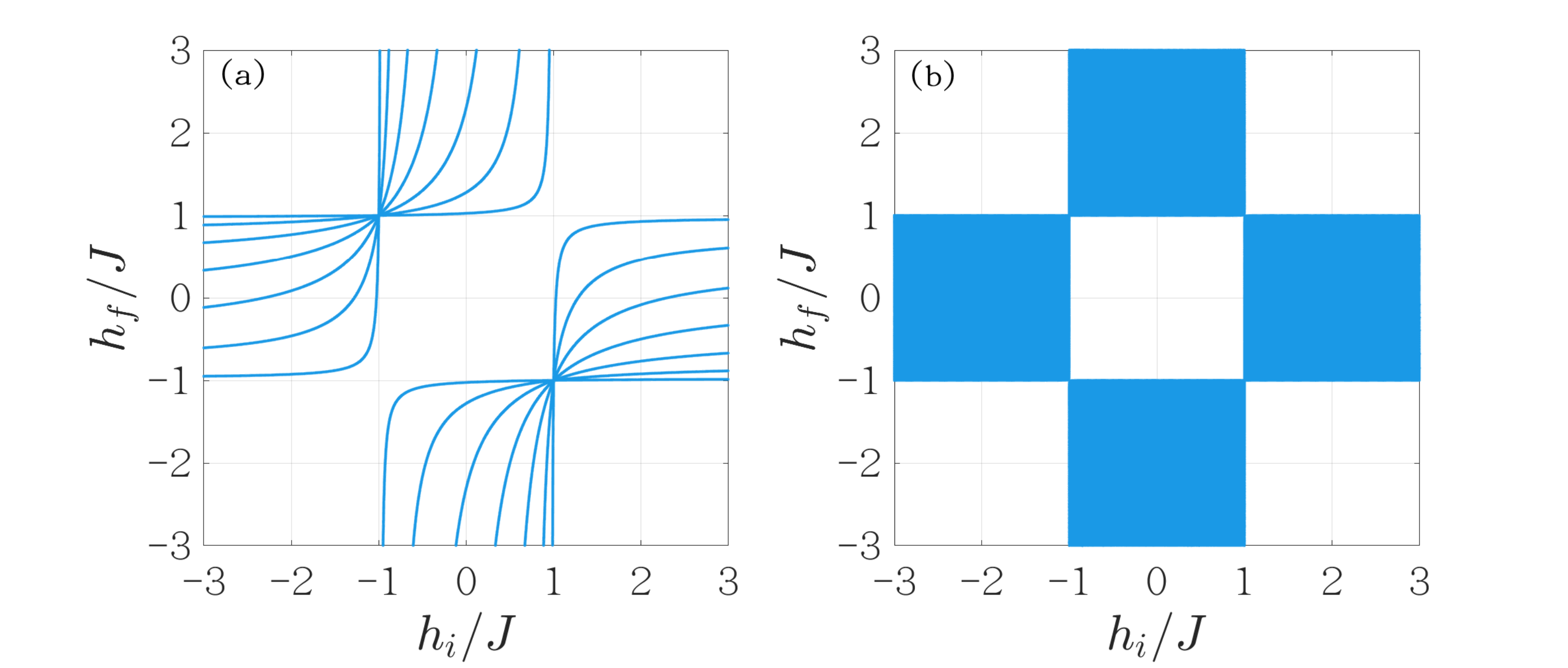}
\par\end{centering}
\centering{}\caption{\label{fig:Ising_hihf_k} The combination of $h_{i}/J$ and $h_{f}/J$
which fulfill Eq. (\ref{eq:Ising_hihf_k}). (a) $L=14$ and (b) $L=400$.}
\end{figure}

As displayed in Fig. \ref{fig:Ising_hihf_k}(a) for the system with
lattice size $L=14$, for a given $h_{i}/J$, only $L/2$ discrete
values of $h_{f}/J$ satisfy Eq. (\ref{eq:Ising_hihf_k}). Continuously
varying $h_{i}/J$ leads to the formation of a series of curves in
the parameter space spanned by $h_{i}/J$ and $h_{f}/J$.
When we increase the lattice
size,  the number of curves increases linearly and the distribution of curves becomes more and more dense, as shown in Fig. \ref{fig:Ising_hihf_k}(b) for the system with
$L=400$. To characterize the average distance between neighboring
curves, we define the quantity $\overline{\Delta}$ as
\begin{equation}
\overline{\Delta}=\frac{1}{L-1}\sum_{k=\frac{1-L}{L}\pi}^{\frac{L-3}{L}\pi}\left|\frac{1}{J}\left[h_{f}(k+\frac{2\pi}{L})-h_{f}(k)\right]\right|,\label{eq:Dbar}
\end{equation}
where $h_{f}(k)$ is the solution of Eq. (\ref{eq:Ising_hihf_k}) for
a $k$ mode. In the thermodynamic limit we can turn the sum into an
integral and it can be found that $\overline{\Delta}$ is approximately
equal to $4/L$ which approaches to $0$ as $L\rightarrow\infty$.
This is also confirmed by the numerical result as displayed in Fig. \ref{fig:Dhf}(a).
Therefore the discrete values of $h_{f}/J$ tend to distribute continuously
in the thermodynamic limit, which is consistent with the general knowledge
about the DQPT that zeros of LE appear when we quench the system
across the critical point.
\begin{figure}[h]
\begin{centering}
\includegraphics[scale=0.21]{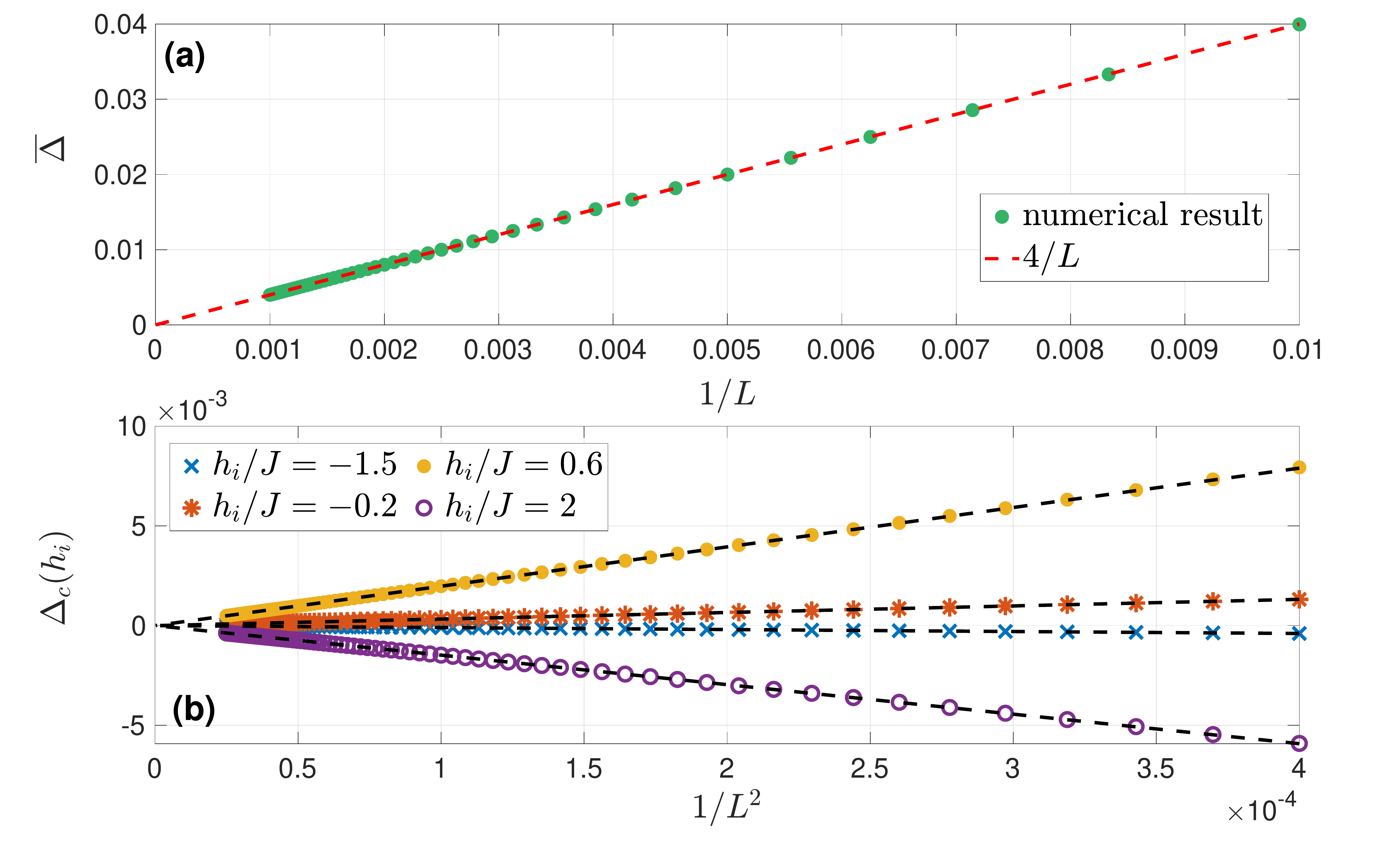}
\par\end{centering}
\centering{}\caption{\label{fig:Dhf} (a) Green dots are the numerical results of Eq. (\ref{eq:Dbar})
for finite size systems. Red dashed line is the result in the large
size limit $\overline{\Delta}\approx4/L$. We set prequench parameter
as $h_{i}/J=1.5$. (b) The $\Delta_{c}(h_{i})$ with respect to $1/L^{2}$
for different values of $h_{i}/J$. The discrete marks are the numerical
results and the dashed lines are the results of Eq. (\ref{eq:Dchi}).}
\end{figure}

If we quench the system to the critical point $h_{f}/J=1$ $(-1)$,
from Eq. (\ref{eq:Ising_hihf_k}), we can see that no exact zeros of
LE are available unless the initial state is prepared in the other
critical point $h_{i}/J=-1$ $(1)$. Interesting, if we restrict $h_{i}/J\geq0$
and $h_{f}/J\geq0$, no exact zero of Loschmidt echo can be found
if the value of $h_{i}/J$ or $h_{f}/J$ is in the interval\textbf{
$\left(-\cos\frac{L-1}{L}\pi,-\sec\frac{L-1}{L}\pi\right)$} for the
finite size system. The interval $\left(-\cos\frac{L-1}{L}\pi,-\sec\frac{L-1}{L}\pi\right)\approx\left(1-\frac{1}{2}\left(\frac{\pi}{L}\right)^{2},1+\frac{1}{2}\left(\frac{\pi}{L}\right)^{2}\right)$
is around the critical point $h_{c}/J=1$ and the boundary of the
interval are reciprocal due to the existence of dynamical duality
for the TFIM. Moreover, we can define a quantity $\Delta_{c}(h_{i})$
which represents the shortest distance between $h_{f}/J=1$ and the
solutions of Eq. (\ref{eq:Ising_hihf_k}) for arbitrary value of $h_{i}/J$.
The numerical result of $\Delta_{c}(h_{i})$ is shown in Fig. \ref{fig:Dhf}(b)
for $h_{i}/J=-1.5,-0.2,0.6,2$ represented by different marks. The
approximate formula of $\Delta_{c}(h_{i})$ for large $L$ can be
derived from Eq. (\ref{eq:Ising_hihf_k}), which reads as
\begin{equation}
\Delta_{c}(h_{i})\approx\frac{\alpha(h_{i})}{L^{2}},\label{eq:Dchi}
\end{equation}
where $\alpha(h_{i})=\frac{\pi^{2}\left(J+h_{i}\right)}{2\left(J-h_{i}\right)}$.
The results of Eq. (\ref{eq:Dchi}) for $h_{i}/J=-1.5,-0.2,0.6,2$
are shown in Fig. (\ref{fig:Dhf})(b) which are denoted by black dashed
lines. So, if we quench the system from arbitrary value of $h_{i}/J$\textcolor{red}{{}
}except $-1$ to $h_{f}/J$ near the critical point, then there exists
a region in which no exact zeros of LE are available for a finite
size system. The width of this region is dependent on $h_{i}/J$ and
$\Delta_{c}(h_{i})\rightarrow0$ in the thermodynamic limit for any
$h_{i}/J$. Together with the result of $\overline{\Delta}\rightarrow0$,
we can see that exact zeros of LE would exist so long as we quench
across the critical point for the infinite size system, in agreement
with the previous work \citep{Heyl2013PRL} in the thermodynamic limit.
The result of the shortest distance between $h_{f}/J=-1$\textcolor{red}{{}
}and the solutions of Eq. (\ref{eq:Ising_hihf_k}) for arbitrary value
of $h_{i}/J$ is similar to Eq. (\ref{eq:Dchi}).

\section{Quantum speed limit time for dynamical quantum phase transition}

From the previous section, we know that there exist exact zeros of
LE as we quench the ground state across the static phase transition
point. It is known that the QSL time is the
minimal time for the evolution of an initial state to its orthogonal
state \citep{Giovannetti2003PRA,Levitin2009PRL,Fogarty2020PRL}, and thus the time for the emergence of the
first exact zero of LE gives the QSL time, i.e.,
\begin{equation}
\tau_{\text{QSL}}=\frac{\pi}{4E_{kf}}.
\end{equation}
According to Eq. (\ref{Ekf}), the QSL time is dependent on the values
$h_{f}/J$.
\begin{figure}[t]
\begin{centering}
\includegraphics[scale=0.225]{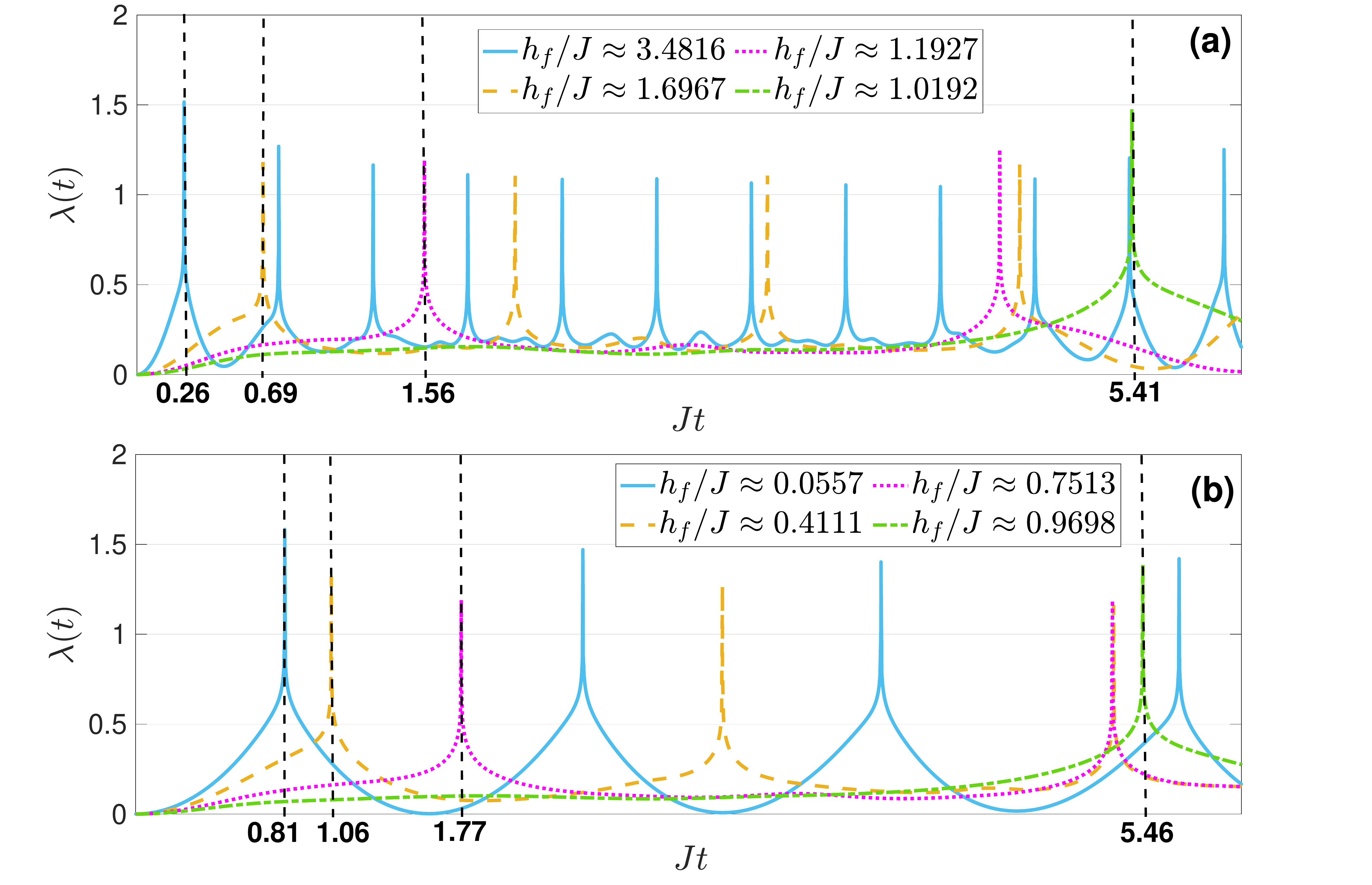}
\par\end{centering}
\centering{}\caption{\label{fig:Los_tmin_L22} Rate function $\lambda(t)$ for $L=22$.
Black dashed lines guide the first time for the appearance of the
exact zero of LE for every $h_{f}/J$. The prequench parameter is
(a) $h_{i}/J=0.3$ and (b) $h_{i}/J=2$.}
\end{figure}

As displayed in Fig. \ref{fig:Los_tmin_L22} for the system with $L=22$,
we show that a series of divergence points of rate function, corresponding
to exact zeros of LE, appear in the real time axis. The QSL time corresponds
to the first divergence point of rate function, which is labeled by
black dashed line. To see the dependence of $\tau_{\text{QSL}}$ on
$h_{f}/J$, we plot rate function versus $Jt$ for all permitted $h_{f}/J>0$
determined by Eq. (\ref{eq:Ising_hihf_k}). It can be observed from the Fig. \ref{fig:Los_tmin_L22}(a) that
the QSL time decreases with the increase in $h_{f}/J$, when we quench from the
initial phase in the region of $0<h_{i}/J<1$ to the region of $h_{f}/J>1$. On the other hand, when we quench from the region of $h_{i}/J>1$ to $0<h_{f}/J<1$, the QSL time decreases with the decrease in $h_{f}/J$ as shown in
Fig. \ref{fig:Los_tmin_L22}(b).
Such an observation does not rely on the system size and can be obtained
from Eq. (7) and Eq. (8). It follows that the
QSL time increases as $h_{f}/J$ approaches the critical
point $h_{f}/J=1$. 

Particularly, we denote
the maximal value of quantum speed limit time as $\tau_{max}=\max\left[\tau_{\text{QSL}}\right]$.
It is found that $\tau_{max}$ is corresponding to the quench process
with the postquench parameter closest to $h_{c}/J=1$.
From the analytical result Eq. (\ref{eq:Ising_hihf_k}) and the formula
of $E_{kf}$, it follows that $E_{kf}/J\approx\frac{\pi}{L}$ is minima
if $k=\frac{\pi}{L}$ or $k=\frac{L-1}{L}\pi$. According to Eq. (\ref{eq:Ising_hihf_k}),
it should also be noted that the mode $k=\frac{\pi}{L}$ and $k=\frac{L-1}{L}\pi$
is corresponding to $h_{f}/J$ closest to $-1$ and $1$ for the finite
size system, respectively. Then we have $\tau_{max}\approx{L}/{(4J)}$
which can be regarded as a upper bound of QSL time. As demonstrated in Fig. \ref{fig:tmax_critical}, we display the QSL time with respect to $L$ for $h_i/J=0.2,0.6,2,10$  and $h_f/J$ taken to be closest to -1 (Fig. \ref{fig:tmax_critical}(a)) and  1 (Fig. \ref{fig:tmax_critical}(b)), respectively.  The red dashed lines in Fig. \ref{fig:tmax_critical} guide the value of $\tau_{\text{QSL}}={L}/{(4J)}$, indicating that  $\tau_{max}$ increases linearly with the increase in the system size. In the thermodynamic
limit, $k\rightarrow0$ and $k\rightarrow\pi$ is corresponding to
$h_{f}/J\rightarrow-1$ and $h_{f}/J\rightarrow1$, respectively.
When $L\rightarrow\infty$, we have $\tau_{max}\rightarrow\infty$
with $\left|h_{f}/J\right|\rightarrow1$. This means that we can not
observe the DQPT in a finite time if we quench the system from a non-critical
phase to the critical phase with $\left|h_{f}/J\right|=1$, i.e., no DQPT occurs in a finite time if we
quench from a non-critical phase to the critical point due
to the corresponding $\tau_{max}$  which is approaching infinity.
%%%%%%%%%%%%%%%%%%%%%%%%%%%%%%%%%%%
\begin{figure}[t]
\begin{centering}
\includegraphics[scale=0.23]{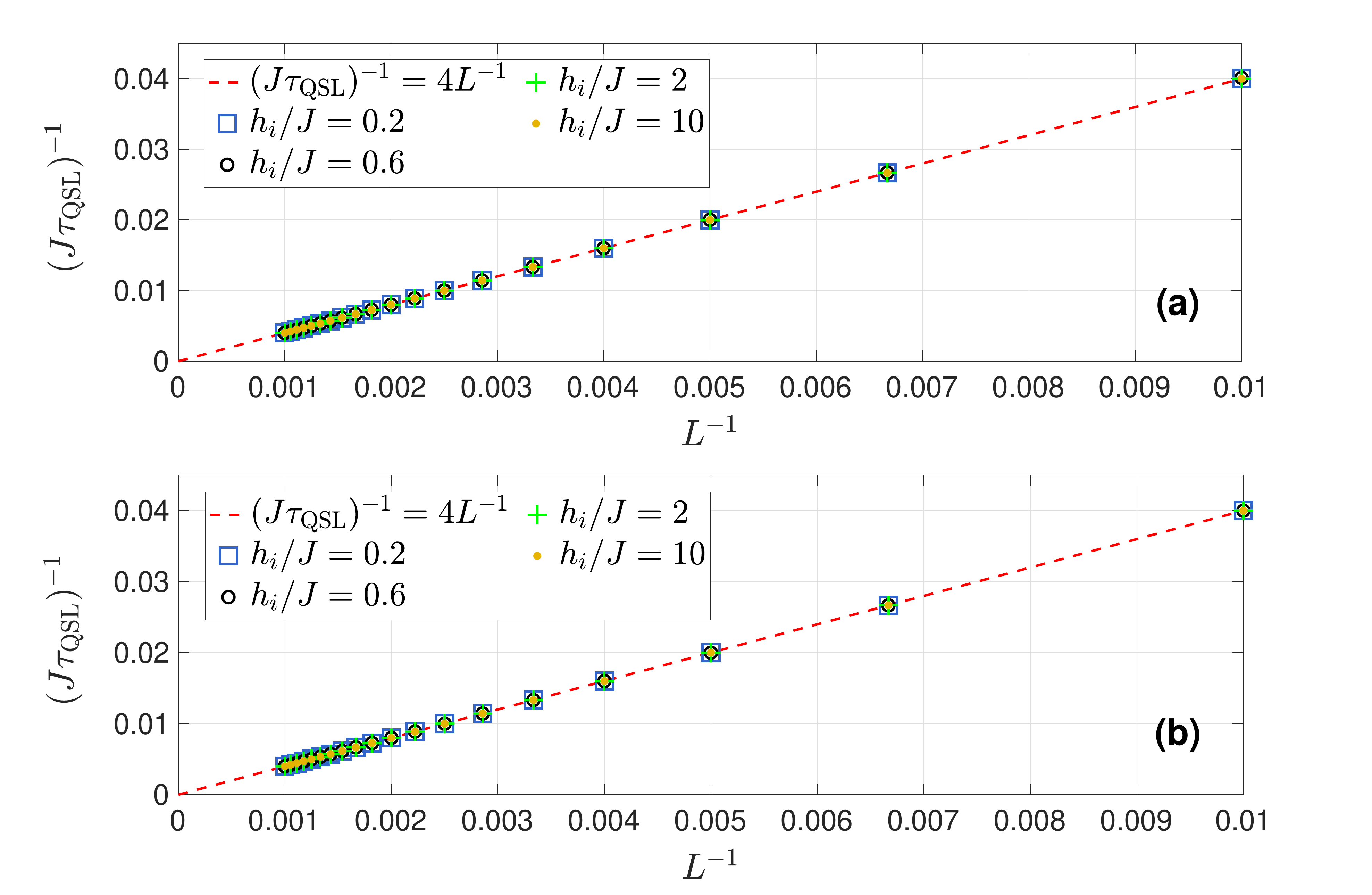}
\par\end{centering}
\centering{}\caption{\label{fig:tmax_critical} $(J\tau_{\text{QSL}})^{-1}$ versus  $L^{-1}$ for different prequench parameters. Here the postquench parameters fulfill Eq. (\ref{eq:Ising_hihf_k}) and are chosen to be closest to (a)  $h_f/J=-1$ and (b) $h_f/J=1$.}
\end{figure}

For any ground state of 1D TFIM, it is also interesting to ask how
fast could the ground state achieve to its orthogonal state as we
quench the parameter to the system? The answer of the question is
given by the minimal value of QSL time denoted as $\tau_{min}=\min\left[\tau_{\text{QSL}}\right]$.
The closely related problem is  Anderson's orthogonality catastrophe which  demonstrates that a local perturbation can cause two many-body states  to achieve orthogonality in the thermodynamic limit. And the connection between Anderson's orthogonality catastrophe and the QSL time in quantum quench dynamics has been disscussed in recent work \citep{Fogarty2020PRL}.
In Fig. \ref{fig:tmin_hi}, we demonstrate the behavior of $J\tau_{min}$
for various prequench parameters $h_{i}/J$ as the
system size increases. It can be found that if the initial state lies
in the paramagnetic phase [Fig. \ref{fig:tmin_hi}(a)], $\tau_{min}$
would approach to some finite values as the system size increases. However, $\tau_{min}$
would approach to  zero if the initial state lies in
the  ferromagnetic phase [Fig. \ref{fig:tmin_hi}(b)].

To get a better understanding, now we discuss two limiting cases.
One is the initial state chosen in the paramagnetic phase, i.e., the
ground state of the prequench Hamiltonian with $h_{i}/J=\infty$.
It can be seen that $\tau_{min}$ is corresponding to the maximal
value of $E_{kf}/J=\sqrt{(\cos k+h_{f}/J)^{2}+\sin^{2}k}$ with $h_{f}/J$
fulfilling Eq. (\ref{eq:Ising_hihf_k}). For $h_{i}/J=\infty$, the
maximal value of $E_{kf}$ in the thermodynamic limit is $E_{kf}/J=1$
corresponding to $k=\pi/2$, so we have $J\tau_{min}=\frac{\pi}{4}$.
We note that the exact zeros of LE in this limiting case have been
discussed in Refs. [5,13,28].
And it can be observed in Fig. \ref{fig:tmin_hi}(a) for $h_{i}/J=1000$,
where the black dashed line guides the value of $\pi/4$. On the contrary,
if we consider the prequench Hamiltonian lying in the ferromagnetic
phase with $h_{i}/J=0$. From Eq. (\ref{eq:Ising_hihf_k}), we have
$h_{f}/J=-1/\cos k$, and it follows that the maximal value of $E_{kf}$
is $E_{kf}=\infty$ corresponding to $k=\pi/2$, which means $\tau_{min}=0$
in the thermodynamic limit. For the finite size system, $k$ can be
exactly equal to $\frac{\pi}{2}$ if $\mod(L,4)=2$, so we have $\tau_{min}=0$.
Otherwise, $k=\frac{\pi}{2}+\frac{\pi}{L}$ is the mode closest to
$\frac{\pi}{2}$ for $\mod(L,4)=0$. In this case, we have $h_{f}/J\approx\frac{L}{\pi}$
for $L\rightarrow\infty$ and the maximal value of $E_{kf}/J\approx\frac{L}{\pi}$
with $k=\frac{\pi}{2}+\frac{\pi}{L}$ . Then we have $J\tau_{\text{QSL }}\approx\frac{\pi^{2}}{4L}$
which is illustrated in Fig. \ref{fig:tmin_hi}(b) by the green solid
line and it is shown that the asymptotic behavior of $\tau_{\text{QSL}}$
is captured by the line of $\frac{\pi^{2}}{4L}$ for $\left|h_{i}/J\right|<1$.
\begin{figure}[h]
\begin{centering}
\includegraphics[scale=0.225]{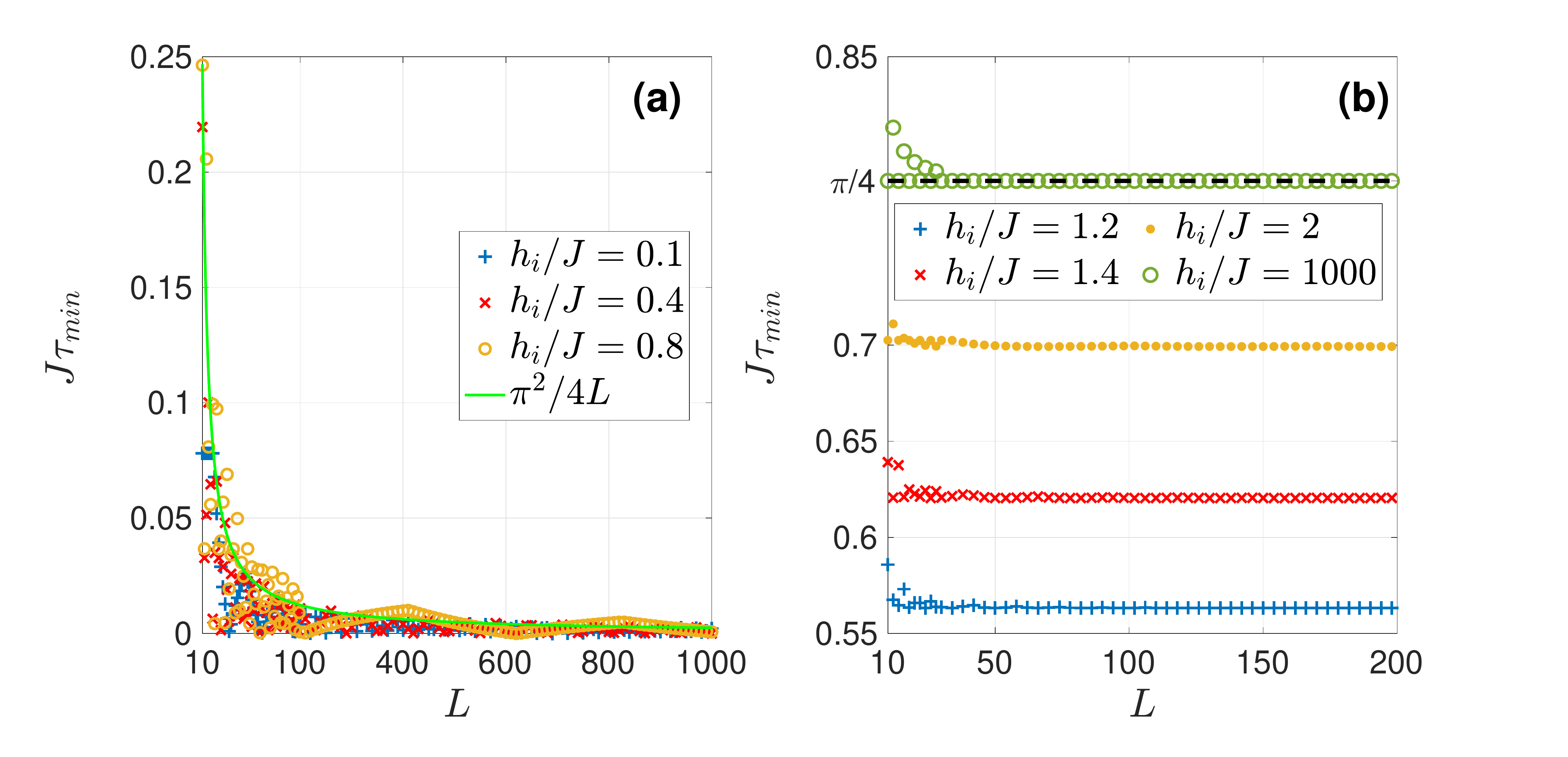}
\par\end{centering}
\centering{}\caption{\label{fig:tmin_hi} $J\tau_{min}$ versus system size $L$. The prequench
parameter is (a) $\left|h_{i}/J\right|>1$ and (b) $\left|h_{i}/J\right|<1$.}
\end{figure}

It is known that the lower bound of QSL time can be given by Mandelstam--Tamm
bound $\tau_{\text{MT}}$ \citep{MT1945JP,Bhat1983JPAMG,Vaidman1992AJP}:
\begin{equation}
\tau_{\text{QSL}}\geq\tau_{\text{MT}}\equiv\frac{\pi}{2\Delta E},\label{eq:tau}
\end{equation}
where $\left(\Delta E\right)^{2}=\langle\psi_{i}|H_{f}^{2}|\psi_{i}\rangle-\left(\langle\psi_{i}|H_{f}|\psi_{i}\rangle\right)^{2}$
with $H_{f}$ denotes the postquench Hamiltonian. Next, we consider
the two cases discussed above. For the case with paramagnetic initial
state $|\psi_{i}\rangle=\otimes_{j=1}^{L}|\uparrow\rangle_{j}$ and
the postquench Hamiltonian\textcolor{red}{{} }$H_{f}=-J\sum_{j=1}^{L}\sigma_{j}^{x}\sigma_{j+1}^{x}$
with $h_{f}/J=0$, we have $\Delta E=J\sqrt{L}$ due to $\langle\psi_{i}|H_{f}^{2}|\psi_{i}\rangle=J^{2}L$
and $\langle\psi_{i}|H_{f}|\psi_{i}\rangle=0$. So the MT
bound is $\tau_{\text{MT}}\rightarrow0$ in the thermodynamical limit.
For the other case with the ferromagnetic initial state $|\psi_{i}\rangle=\otimes_{j=1}^{L}|\rightarrow\rangle_{j}$
(or $|\psi_{i}\rangle=\otimes_{j=1}^{L}|\leftarrow\rangle_{j}$) and
$H_{f}=-h_{f}\sum_{j=1}^{L}\sigma_{j}^{z}$ with $h_{f}/J\rightarrow\infty$,
we have $\Delta E\rightarrow\infty$ and $\tau_{\text{MT}}\rightarrow0$.
It can be seen that the MT bound of QSL time is equal
to zero for both two cases. In comparison with our exact result of
$\tau_{\text{QSL}}$, it can be found that the MT bound
$\tau_{\text{MT}}$ is tight if the prequench Hamiltonian lies in
the ferromagnetic phase.

To see clearly how $\tau_{min}(L)$ changes with $h_{i}/J$, we can
calculate the mean value of $\tau_{min}(L)$ numerically from $L_{min}$
to $L_{max}$ and denote it as:
\begin{equation}
\bar{\tau}_{min}=\frac{1}{L_{max}-L_{min}}\sum_{L=L_{min}}^{L_{max}}\tau_{min}(L).\label{eq:meant_L}
\end{equation}
Meanwhile, we can also calculate the variance of $\tau_{min}(L)$
defined by:
\begin{equation}
\sigma_{\tau_{min}}^{2}=\frac{1}{L_{max}-L_{min}}\sum_{L=L_{min}}^{L_{max}}\left[\tau_{min}^{2}(L)-\bar{\tau}_{min}^{2}\right].\label{eq:vart_L}
\end{equation}
We count from $L_{min}=10$ to $L_{max}=10000$ and show the numerical
results of $ J\bar{\tau}_{min}$ and $J^2\sigma_{\tau_{min}}^{2}$ with
respect to prequench parameter $h_{i}/J$ in Figs. \ref{fig:MV_tmin}(a)
and 6(b), respectively. It can be observed that both $\bar{\tau}_{min}$
and $\sigma_{\tau_{min}}^{2}$ have an abrupt change at $h_{i}/J=1$
which corresponds to the static quantum phase transition point in
the thermodynamic limit. The fluctuation of energy can be evidenced
in $\sigma_{\tau_{min}}^{2}$ (Fig. \ref{fig:MV_tmin}(b)) which remains
a nonzero value as $\left|h_{i}/J\right|<1$ and diverges as $\left|h_{i}/J\right|$
approaches $1$. The non-analytical behaviours appearing in the change
of prequench parameter across the static quantum phase transition
point indicates that clearly the minima of QSL time relies on the choice
of initial states.
%%%%%%%%%%%%%%%%%%%%%%%%%%%%%%%%%
\begin{figure}[h]
\begin{centering}
\includegraphics[scale=0.22]{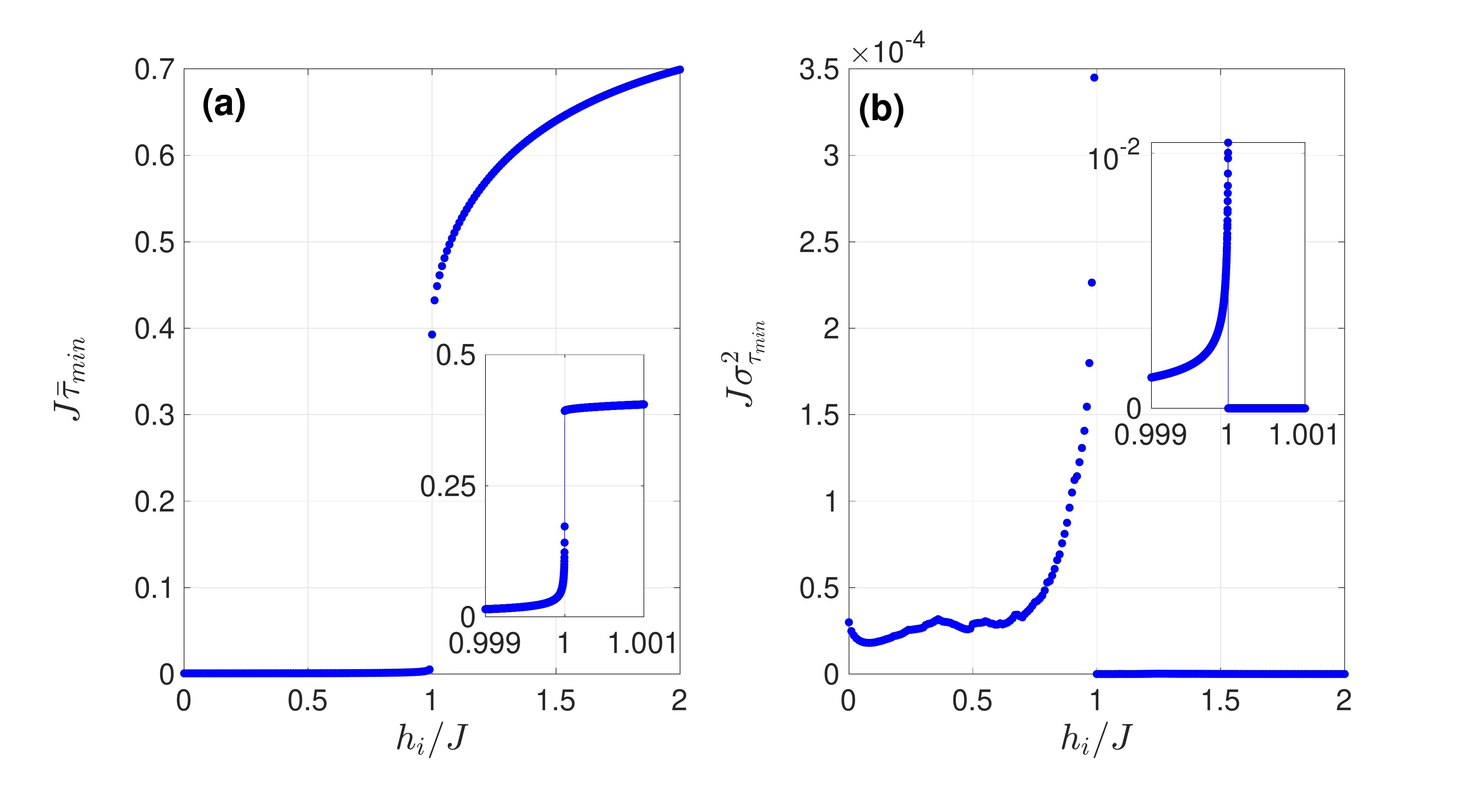}
\par\end{centering}
\centering{}\caption{\label{fig:MV_tmin} (a) $J \bar{\tau}_{min}$ with respect to $h_{i}/J$;
(b) $J^2 \sigma_{\tau_{min}}^{2}$ with respect to $h_{i}/J$. Here we
count the size of system from $L_{min}=10$ to $L_{max}=10000$.}
\end{figure}

\section{Summary and outlook}

In summary, we have analytically calculated the exact zeros of the
LE for the 1D TFIM and shown that there exist exact zeros of LE even
for the finite size quantum system when the post-quench parameter
takes some discrete values. As the system size tends to infinity,
the discrete parameters distribute continuously in the parameter regions
with the corresponding equilibrium phase different from the initial
phase, which is in agreement with previous work in the thermodynamic
limit. We also unveil that no exact zeros of LE are available if we
quench the system to (or from) the critical point. For the finite size
systems of 1D TFIM, we have unveiled how the QSL time changes
with quench parameters and
studied the behaviors of the maximum and minimum values of the QSL time. From our analytical result in the thermodynamic limit,
it is shown that no DQPT occurs in a finite time if we quench from a
non-critical phase to the critical point due to that the corresponding $\tau_{max}$
is approaching infinity. We have also illustrated the existence of
nonanalytical behaviors in both the average of $\tau_{min}(L)$ and
the variance of $\tau_{min}(L)$ when we change the parameter of prequench
Hamiltonian across the underlying static critical point.

Our work provides a firm theoretical ground for understanding why
and how the DQPT occurs with the increase of system sizes and the
peculiar dynamical behavior near the critical point, which paves the
way for experimental investigations of DQPT for small size systems.
According to our theoretical finding, we can always find exact zeros
of LE by tuning the quench parameter to a series of discrete fine-tuning
points for the finite size system which supports DQPT in the thermodynamical
limit. At these fine-tuning points, the divergence of the corresponding
rate function can be observed in some critical times. The number of
fine-tuning points increases linearly with the increase of lattice
size. By recording the critical times for the emergence of exact zeros
of LE, one can also experimentally study the behaviors of quantum
speed limit time and explore its connection to the DQPT.

\begin{acknowledgments}
This work is supported by National Key Research and Development Program
of China (2016YFA0300600), NSFC under Grants No.11974413 and the Strategic
Priority Research Program of Chinese Academy of Sciences under Grant
No. XDB33000000.
\end{acknowledgments}

\appendix

\section{The dynamical duality relation of the Loschmidt echo}
Consider the quench dynamics of the quantum TFIM by suddenly switching the
transverse field from the prequench parameter $h_{i}$ to the postquench parameter $h_{f}$.  The Loschmidt echo can be represented as
\begin{equation}
\mathcal{L}(\gamma_{i},\gamma_{f},t)=\prod_{k}\mathcal{L}_{k}(\gamma_{i},\gamma_{f},t),
\end{equation}
where $\gamma_{i}=h_{i}/J$ and $\gamma_{f}=h_{f}/J$ are the dimensionless
parameters and the $k-$component of the Loschmidt echo is
\begin{equation}
\mathcal{L}_{k}(\gamma_{i},\gamma_{f},t)=1-\sin^{2}\left(2\delta\theta_{k}\right)\sin^{2}\left(2E_{kf}t\right).
\end{equation}
Here we have
\begin{align*}
 & \sin^{2}\left(2\delta\theta_{k}\right)\\
= & \left(\frac{\zeta_{f}}{E_{f}}\frac{\epsilon_{i}}{E_{i}}-\frac{\epsilon_{f}}{E_{f}}\frac{\zeta_{i}}{E_{i}}\right)^{2}\\
= & \frac{\left(\zeta_{f}\epsilon_{i}-\epsilon_{f}\zeta_{i}\right)^{2}}{\left(\epsilon_{i}^{2}+\zeta_{i}^{2}\right)\left(\epsilon_{f}^{2}+\zeta_{f}^{2}\right)}\\
= & \frac{\left(\gamma_{i}-\gamma_{f}\right)^{2}\sin^{2}k}{\left(1+2\gamma_{i}\cos k+\gamma_{i}^{2}\right)\left(1+2\gamma_{f}\cos k+\gamma_{f}^{2}\right)},
\end{align*}
and
\begin{align*}
\sin^{2}\left(2E_{kf}t\right) & =\sin^{2}\left(2J_{f}t\sqrt{1+2\gamma_{f}\cos k+\gamma_{f}^{2}}\right)
\end{align*}

Now we consider the case with the prequench and postquench dimensionless
parameters being $1/\gamma_{i}$ and $1/\gamma_{f}$, respectively. The $k-$component
of Loschmidt echo of the corresponding model can be written as
\begin{equation}
\mathcal{L}_{k}(\frac{1}{\gamma_{i}},\frac{1}{\gamma_{f}},t)=1-\sin^{2}\left(2\delta\tilde{\theta}_{k}\right)\sin^{2}\left(2\tilde{E}_{kf}t\right),
\end{equation}
with
\begin{align*}
 & \sin^{2}\left(2\delta\tilde{\theta}_{k}\right)\\
= & \frac{\left(\gamma_{i}^{-1}-\gamma_{f}^{-1}\right)^{2}\sin^{2}k}{\left(1+2\gamma_{i}^{-1}\cos k+\gamma_{i}^{-2}\right)\left(1+2\gamma_{f}^{-1}\cos k+\gamma_{f}^{-2}\right)}\\
= & \frac{\left(\gamma_{i}-\gamma_{f}\right)^{2}\sin^{2}k}{\left(\gamma_{i}^{2}+2\gamma_{i}\cos k+1\right)\left(\gamma_{f}^{2}+2\gamma_{f}\cos k+1\right)}\\
= & \sin^{2}\left(2\delta\theta_{k}\right),
\end{align*}
and
\begin{align*}
\sin^{2}\left(2\tilde{E}_{kf}t\right)=\sin^{2}\left(2E_{kf}\gamma_{f}^{-1}t\right)
\end{align*}
where $\tilde{E}_{kf} \equiv E_{kf}(1/\gamma_{f})$ and we have used the dual relation of eigenvalues $E_{kf}(\gamma_{f})=\gamma_{f} E_{kf}(1/\gamma_{f})$.

Then we can observe that
\begin{equation}
\mathcal{L}_{k}(\gamma_{i},\gamma_{f},t)=\mathcal{L}_{k}(\gamma_{i}^{-1},\gamma_{f}^{-1},\gamma_{f}t).
\end{equation}
which gives rise to the dynamical dual relation Eq. (\ref{eq:ddr})
directly.

\section{The case under periodical boundary condition}

\begin{figure}[h]
\begin{centering}
\includegraphics[scale=0.23]{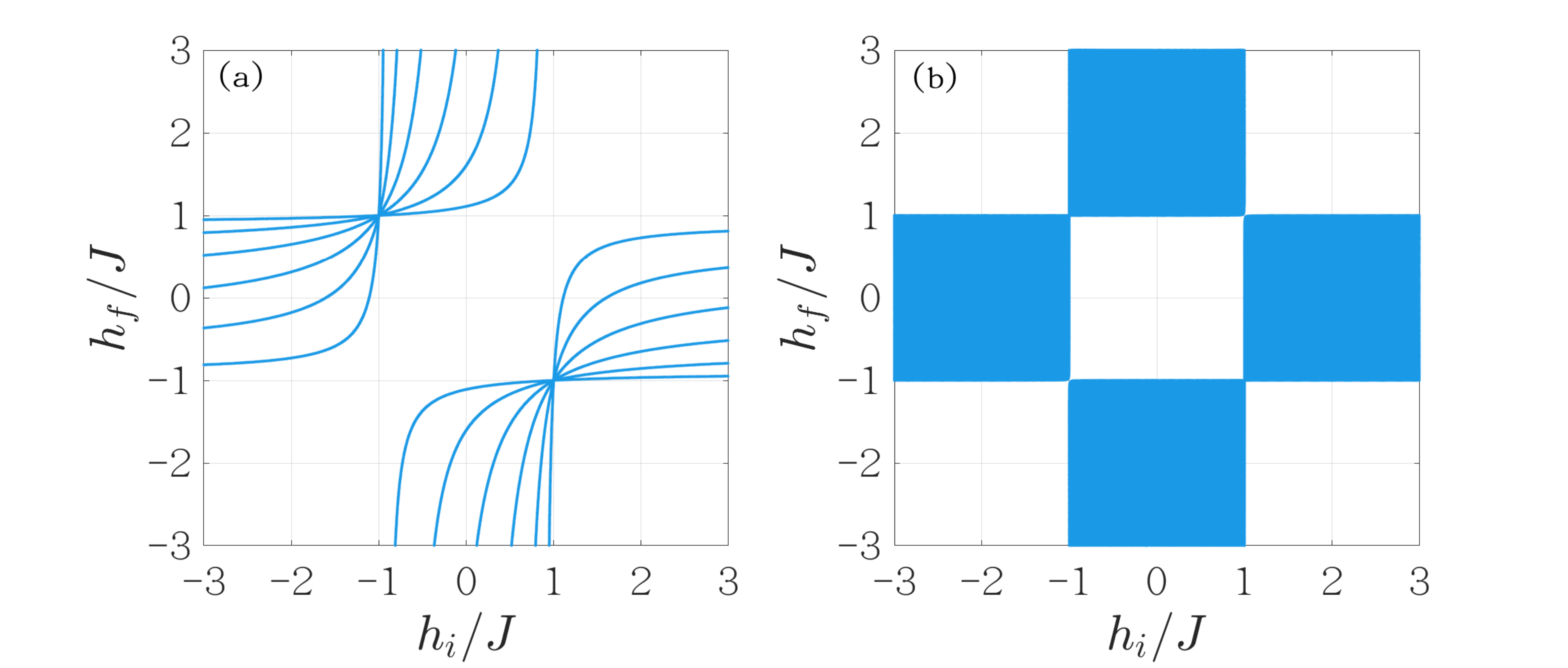}
\par\end{centering}
\centering{}\caption{\label{fig:Ising_hihf_k_PBC} The combination of $h_{i}/J$ and $h_{f}/J$
which fulfill Eq. (\ref{eq:Ising_hihf_k}) under the PBC for odd parity
space. (a) $L=14$ and (b) $L=400$.}
\end{figure}

In the main text, we have taken the anti-periodical boundary condition.
In this appendix, we consider the odd parity of the 1D TFIM which
corresponds to the periodical boundary condition of Hamiltonian Eq. (\ref{H-JW}).
The formulas for determining exact zeros of LE in the main text do
not change, and the constraint relation of quench parameter is the
same as the Eq. (\ref{eq:Ising_hihf_k}). However, the values of $k$
under PBC should be chosen in the set  of $\mathcal{K}_{\text{PBC}}=\left\{ k=\frac{2\pi m}{L}|m=-L/2+1,\cdots,0,\cdots,L/2\right\} $.
Here the modes corresponding to $k=0$ and $k=\pi$ should be removed
due to $\sin^{2}(2\delta\theta_{k})=0$ and thus $\mathcal{L}_{k}=1$
for $k=0$ or $k=\pi$. Therefore, under the PBC, for a given $h_{i}/J$,
only $L/2-1$ discrete values of $h_{f}/J$ satisfy Eq. (\ref{eq:Ising_hihf_k})
corresponding to $k=\frac{2\pi}{L},\cdots,\frac{2\pi(L/2-1)}{L}$.
We display the result of Eq. (\ref{eq:Ising_hihf_k}) in Fig. \ref{fig:Ising_hihf_k_PBC}(a) and (b) for $L=14$ and $L=400$ under the PBC, respectively,
in contrast to Fig. 1 for the same system under the aPBC. When the
system size tends to infinity, the discrete values of $h_{f}/J$ tend
to distribute continuously, and therefore our conclusions do not rely
on boundary conditions in the thermodynamical limit.

\end{document}